**Ab initio computational modeling of tumor spheroids**

Roberto Chignola[1,2], Alessio Del Fabbro[2], Marcello Farina[3], Edoardo Milotti[2,4]

[1]*Dipartimento di Biotecnologie, Università di Verona, Strada le Grazie 15 - CV1, I-37134, Verona, Italia*

[2]*Istituto Nazionale di Fisica Nucleare, Sezione di Trieste, Via Valerio 2, I-34127, Trieste, Italia*

[3]*Dipartimento di Elettronica e Informazione, Politecnico di Milano, Via Ponzio 34/5, I-20133, Milano, Italia*

[4]*Dipartimento di Fisica, Università di Trieste, Via Valerio 2, I-34127, Trieste, Italia*

# 1. Introduction

*1.1 Tumor growth as a multi-scale biological process*

Biological systems span vast spatiotemporal scales, from the microscopic dynamics of atoms to the macroscopic dynamics of cell clusters. Information flows in both directions and determines the behavior of living matter and ultimately the normal physiology of organisms and the onset of pathologies such as tumors. Individual tumors are complex biological systems and, in spite of great therapeutic advances, many tumors still escape treatment and lead to death. Indeed, the malignancy and the response of tumors to therapy depends on their growth potential which in turn is determined by the ability of tumor cells to adapt to different environments, to compete with normal cells for both space and nutrients and to ignore molecular signals attempting to block cell cycling or to promote cell death (Hanahan and Weinberg 2000). Major efforts have been made by experimenters to highlight the molecular circuits underlying tumor cell biology and, on this basis, to develop novel therapeutic strategies. This has resulted in a huge body of knowledge which has deepened our understanding of the molecular details of tumor cell biology but often with little or no consequence on the clinical management of tumors.

At least part of the complexity of the problem is a sheer consequence of tumor size: clinicians deal with the macroscopic properties of tumors, i.e. masses that may eventually weigh a few kilograms, and thus with a number of cells that ranges between $10^6$ and $10^{13}$, and that may grow for months or years, with a corresponding number of cell cycles somewhere in the range between 100 and 10000. Moreover, at the microscopic level the malignant transformation of single cells is a multistep process that involves the modification of several molecular circuits which, in turn, modify the cell's behavior and the relationships between cells and the environment (Hanahan and Weinberg 2000). In addition, epigenetic and environmental factors, which include cell-cell interactions, also conspire with genetic information to make tumor growth a highly variable process with very strong feedbacks (Hanahan and Weinberg 2000). The highly nonlinear character of the cells' internal molecular machinery, combined with the cell-cell and environmental interactions, with the large number of cells in a tumor, and with the extended tumor lifespan, make predictions based on the behavior of a single molecular circuit quite haphazard.

The availability of powerful computers has already helped bridge the gap between observations and predictions in many complex problems, and suggest that in the future we shall be able to simulate the behavior of large cell populations *ab initio*, starting from individual molecular reactions in single cells and climbing the ladder of complexity up to the behavior of whole multicellular organisms. To reach this goal, however, we think that it is fundamental to proceed in an incremental way by developing quantitative biophysical models of the cell with increasing complexity and bearing in mind that each modeling step must be validated by experimental observations.

*1.2 Multicell tumor spheroids: an in vitro cell model with intermediate complexity between real tumors and conventional tumor cell cultures*

The main problem when managing solid tumors in the clinical practice does not lie in the treatment of large tumor masses, which are usually removed by surgery, but rather in the control of small masses which are near or below the limits of imaging diagnostics (about 1 mm$^3$). These small tumor cell aggregates may escape conventional treatment and, in time, may lead to recurrence of the primary pathology, often with a different phenotype (e.g., acquired resistance to chemotherapy, acquired ability to metastatize, etc.) (Köstler et al. 2000; Böckmann et al. 2001); cells can also grow up to masses of this size without the support of the vasculature, although recent work is challenging this traditional view (see Vajkoczy et al. 2002, and references cited therein).

Unfortunately it is very difficult to study these micromasses in humans as well as in animal models, because their size is below the imaging limit and it is not possible to measure their biological parameters and thus obtain information to validate the results of numerical simulations. Multicellular tumor spheroids represent a valid and effective experimental cell culture technique capable of preserving the three-dimensional topology of actual tumor cell clusters (Sutherland 1988; Mueller-Klieser 1997). Indeed, it is the three-dimensional topology that determines many important biological features, like the expression of specific genes, a slowed-down diffusion of nutrients and waste, and also the expression of new phenotypes like the resistance to radiotherapy, and in fact multicell tumor spheroids display many interesting biological properties that cannot be observed in monolayer cultures such as (Sutherland 1988; Mueller-Klieser 1997):

1. heterogeneous expression of membrane receptors (that regulate cell adhesion and metabolism and also may act as target for specific anti-tumor drugs);

2. production of an intercellular matrix (important for cell aggregation and for penetration of cells of the immune system);

3. heterogeneous distribution of nutrients and oxygen that lead to the formation of a necrotic core and to a gradient of cell proliferation;

4. appearance of resistance phenomena and/or heterogeneous response to antitumor therapies;

5. growth kinetics very similar to those observed *in vivo*.

Multicell tumor spheroids are thus intermediate between traditional cell cultures and tumor *in vivo*, and at the same time they are accessible to experimental measurements: they provide many data that can be used to test and validate multi-scale models of solid tumor growth in the prevascular phase. They are morphologically similar to small tumors below the detection threshold, and they share with them the lack of vascularization. For these reasons tumor spheroids are the perfect targets for a first computational model of tumor growth.

*1.3 Ab initio mathematical modeling: what does it mean?*

The definition of the term *ab initio* in the Oxford American dictionary is "from the beginning". The question, therefore, is: from which beginning should one start modeling tumor cell proliferation? In principle, one might think of the atoms in the cell: using the methods of molecular dynamics it would then be possible to simulate a living being starting from atoms, molecules, and a description of the forces that bind them (see e.g. Phillips et al. 2002). Unfortunately, at present it is unthinkable to simulate anything more complex than very small objects and for a very short time span, much less than the actual spatiotemporal scales of real tumors and even tumor spheroids. We can exemplify the complexity issue by considering here the problem of exploding memory size. If we take a cell radius of 5 μm, then the cell volume is approximately $5 \times 10^{-16}$ m$^3$, and the corresponding cell mass is about $5 \times 10^{-10}$ g, and this means that a single cell corresponds to about $10^{13}$-$10^{14}$ atoms. On the other hand, if we take a spatial resolution 0.01 nm (approximately one tenth of the diameter of a hydrogen atom), and aim to simulate a system size of 1 mm, then for each coordinate we need a 24-bit dynamic range (3 bytes per coordinate), and thus at least 19 bytes/atom (3 coordinates plus 3 velocities plus one atom label), and about $10^{14}$-$10^{15}$ bytes/cell. Finally, the full simulation of a $10^6$ cell spheroid would require at least $10^{20}$-$10^{21}$ bytes/spheroid, and we see that present-day computers are pitifully inadequate for such a brute force approach. Likewise, the time complexity of simulation algorithms is also unmanageable: here we must assume that we are somehow able to tame the $O(N^2)$ complexity of binary interactions between the simulation elements and also the complexities of several subalgorithms like matrix inversion and the like, and that the overall algorithmic time-complexity of the simulation program adds up to a mere $O(N)$. Since the fastest dynamics in acqueous solutions is determined by the motion of protons in the hydrogen bonds in water, and is of the order of 1 ps (Agmon 1995), and since there are approximately $10^{13}$ hydrogen bonds in each cell, then one must take at least $10^{25}$ time steps just to simulate 1 s of proton motion in a cell (and with a rather poor time resolution), and $\approx 10^{31}$ steps to do the same for a one-million cell spheroid.

These approximate calculations amount to an operational definition of biological complexity, and they show that at the moment we cannot even dream to carry out a true *ab initio* simulation of tumor spheroids. It is thus quite clear that we cannot start developing a model of tumor spheroid from the atomic scale, and yet, we would like to model the individual cells' behaviors that determine many interesting biological properties of the spheroid itself. We must therefore choose a mesoscopic scale somewhere in between, although the question remains whether it is possible to stay at this level and still be able to describe a cell behavior which is in turn determined by the intracellular molecular machinery.

## 2. Models

*2.1 A minimal model of the tumor cell*

When viewed at the mesoscopic scale the intracellular molecular machinery appears to possess an astonishing complexity, with a huge number of intertwined chemical reactions that mark the different phases of the cell's life. On the other hand, a biophysical simulator of tumor spheroids must start from a realistic description of the tumor cell, and this ultimately means that at least cell metabolism and its interconnection with the cell cycle must be modeled at a sufficient level of detail in order to describe how cell behavior is affected by the other cells in the cluster, by the chemical composition of the environment and by physical parameters such as temperature, density, radiation, etc.

Our approach is based on the fact that biochemical networks in the cell possess a hierarchical structure (Barabasi and Oltvai 2004), and if a network has such a topology then the system dynamics is known to be dominated by the network's hubs (Barabasi and Oltvai 2004). Thus, by modeling the hubs of the cell's biochemical network one should, at least in principle, be able to capture most of the information of the cell dynamics. In our mesoscopic approach several details of cell metabolism and of the cell cycle have been parameterized and averaged (for details, see Chignola and Milotti 2005; Chignola et al. 2007). In a sense, we try to apply to cell biology the methods that have been so succesful in statistical mechanics, and set up a kind of *statistical cell biology*, i.e., we neglect many fine details and study the cell cluster as a whole, much like a physicist studies magnetic phase transitions with the rather crude Ising or Heisenberg models (Binney et al. 1998). As a consequence, we achieve a huge reduction in computational complexity and a considerable reduction of the space-time scale problems that affect simulations aimed at calculating the properties of macroscopic objects starting from microscopic models.

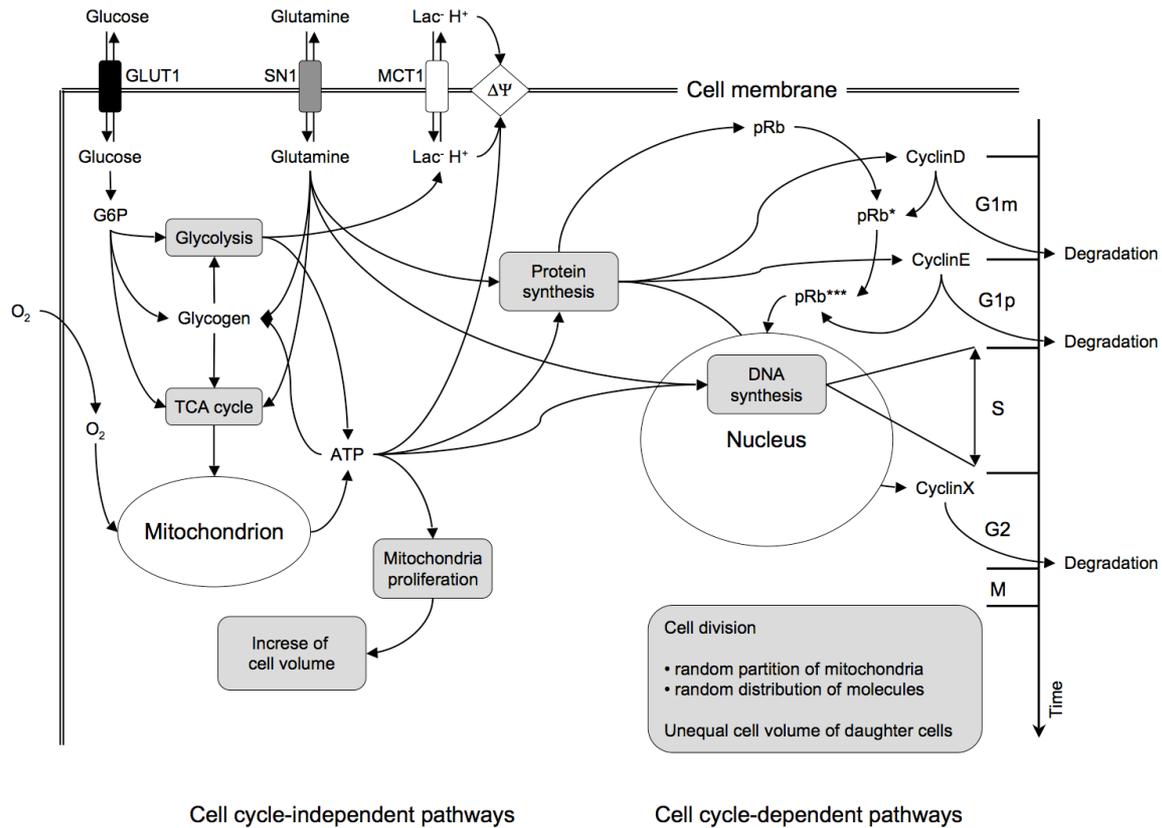

Fig.1 - Metabolic network implemented in the simulation program. See text for details. A complete description of this network can be found in (Chignola et al. 2007).

Figure 1 shows a sketch of the metabolic pathways that have been modeled so far (see Chignola and Milotti 2005; Chignola et al. 2007). We have taken into consideration both cell cycle-independent and cell cycle-dependent pathways. Among the former are the uptake and utilization of nutrients such as glucose and glutamine and the export out of the cell of waste molecules such as lactate. This part describes quantitatively the production of ATP by both oxydative phosphorylation and glycolysis and thus provides a mean to bridge together the two paths: in fact, ATP and glutamine constitute the building blocks of proteins and DNA (Levintow et al. 1955; Salzman et al. 1958; Nelson and Cox 2005) and their availability drives the kinetics of protein and DNA synthesis. These in turn determine the expression of proteins that regulate the cell cycle such as pRb and cyclins and the duration of specific phases of the cell cycle such as the S phase (Weinberg 1995). Proteins that regulate the cell cycle also determine the duration of the other phases, and we have implemented a thresholding mechanism based on multisite phosphorylation (Chignola et al. 2006; Milotti et al. 2007) to determine when a cell steps beyond cell cycle checkpoints.

ATP availability also determines the proliferation of mitochondria, a biological process that has been coupled phenomenologically to the increase of the cell volume, while ATP deprivation leads to cell death (although our model also takes into account several other mechanisms of cell death. See Chignola et al. 2007 for details). Finally, we take into consideration stochastic aspects such as the random partitioning of mitochondria and the random distribution of molecules at cell division, that contribute to desynchronize the duration of the cell cycle in daughter cells (Chignola and Milotti 2005; Chignola et al. 2007). On the whole our numerical model includes continuous deterministic processes (e.g. glucose uptake and utilization), stochastic processes and discrete events (e.g. initiation and termination of DNA duplication and cell division), and its mixed nature leads to serious stability issues in the numerical integration of differential equations (see below).

*2.2 Modeling tumor spheroids*

The minimal model of the tumor cells shown in figure 1 is central to our simulation program and, for validation purposes with experimental data, it has been used to simulate the growth of large cell populations in both closed and open environments (see Chignola and Milotti 2005; Chignola et al. 2007 and the Results section). To model a tumor spheroid, however, we must consider several other biological, chimical and physical aspects as well (Chignola and Milotti 2004; Milotti et al. 2008a):

*1. diffusion of chemical species*. The diffusion of chemicals in a cell cluster proceeds either by normal diffusion or by facilitated diffusion across cell membranes. Facilitated diffusion is mostly a biochemical process that has a weaker dependence on concentration gradients and is brought about essentially by transporters expressed at the cell surface. Obvioulsy, neighboring cells contribute to modify the surrounding environment and eventually compete for the same resources, and thus it is important to define the proximity relationships among cells in the cluster. This is accomplished by a specialized part of the program (see below);

*2. environment*. The enviroment is also taken into account in the simulation: it is defined as the external volume of nutrient fluid that communicates by diffusion with the extracellular spaces that surround cells. The enviroment is modified both by the fast diffusion processes that transport nutrients and metabolites into and out of the cell cluster, and by the slow flushing of nutrient fluid in a bioreactor enclosure;

*3. biomechanical interaction*. In a cluster, cells interact mechanically as well as biochemically: this part of the simulation program is essentially a simple integrator like those found in dissipative dynamics simulations. Cells are approximated by soft spheres that move in a highly viscous environment (see also Chignola and Milotti 2004);

*4. geometry*. Both the mechanical and the diffusion part of the program require a knowledge of the proximity relationships among neighboring cells. We wish to point out that, unlike other models, in our simulation program cells do not grow in an environment defined by a fixed grid, where it is difficult to model appropriately processes such as cell division, and that cells are free to move in the three-dimensional space as as they are pushed and pulled by forces resulting from biomechanical interactions between cells. The nearest neighbors are defined by the links in a Delaunay triangulation (De Berg et al. 2000) and they are computed by the triangulation methods in the computational geometry package CGAL (see

http://www.cgal.org). In this way all the computational complexity of binary interactions is reduced from a potential $O(N^2)$ to a much more manageable $O(N)$.

*2.3 Stability issues in the numerical integration of model differential equations*

In our model each cell is described by a reduced metabolic network and by other mechanisms that include both discrete deterministic and stochastic events (see above). The description is thus mixed, with smooth evolutions interspersed with discrete steps. The exchange of molecules with the surrounding environment means that transport into and out of cells is closely linked with diffusion processes that involve the whole cluster of cells, and finally lead to a very large set of (time) differential equations. Since our goal is to simulate tumor spheroids up to a diameter of 1 mm, which corresponds to about 1 million cells, the software must eventually solve a very large number of coupled nonlinear equations, as many as $10^7$-$10^8$ equations (because there shall be about 10-100 variables per cell). These equations are similar to other equations encountered in systems biology, but the number is uncommonly large. In addition, the equations are quite stiff, since they describe processes that range from fast diffusion in small extracellular spaces (approximately $10^{-6}$ s) up to the slow development of the spheroid as a whole (approximately $10^7$ s) and thus the characteristic times span about 12 orders of magnitude. We have solved the complex stability problems that arise in such a situation (Milotti et al. 2009) and thus our model now stands up as a true multiscale model, both in space and, even more so, in time.

We also remark that some of the model parameters slowly change as cells grow and this is once again at variance with most differential systems used in systems biology where parameters are fixed. Finally, the continuous system evolution described by the differential system is interrupted at random times by discrete events; these events may be internal transition in individual cells (in this case the system parameters change abruptly from one integration step to the next) or cellular division events (in this case the number of equations changes). For all these reasons, we believe that our simulation program is unique in the sense that it tackles simultaneously for the first time a vast array of technical issues that are not addressed by any other model. In particular, the same implementation scheme of model equations can be used to simulate both fast and slow processes simply by tuning the integration time step, and this is an added value to our simulation program that might be exploited to investigate at will biological events with different characteristic times.

## 3. Results

*3.1 Simulations of the growth of dispersed cells in a closed environment*

One important aspect of our approach is that we test the models with actual experimental data: we do not ask the models simply to provide outputs in qualitative agreement with observations, but we require simulation outputs that are quantitatively comparable to real data.

The minimal model of cell metabolism, growth and proliferation described above can be used to simulate a population of dispersed cells growing in a closed environment. This is equivalent to considering cultures of blood cells *in vitro*, such as leukemia cells. Cells growing in a closed environment establish a sort of negative feedback with the environment itself. While real and simulated cells grow, in fact, they consume nutrients and release waste molecules that acidify the medium. As the environment gradually becomes more and more acidic, the uptake of nutrients is also reduced and can eventually switch off completely, thereby leading to a depletion of the energy reserves and ultimately to cell death. This mechanism involves the whole model of cell metabolism and control of the cell cycle and can be tested experimentally because it defines the carrying capacity of the environment where cells are grown.

Our model nicely reproduces common growth curves observed in vitro (Chignola et al. 2007) and shows predictive capabilities, in that it can also describe the growth of leukemia cells under non conventional conditions, such as in growth media whose biochemical composition is periodically modified, that were not considered during model development (Chignola et al. 2007). This is an important test, because it demonstrates that our minimal model of the cell is not just a qualitative description, but is a truly predictive model, and that it can be used as a true *in silico* laboratory.

Indeed, the model provides outputs that are in good quantitative agreement with actual data (Chignola and Milotti 2005; Chignola et al. 2007) on metabolic and cellular parameters (see Table 1).

*3.2 Simulations of the growth of dispersed cells in an open environment*

The negative feedback between cells and their environment discussed above can be partially removed by opening up the environment. In this case, there is continuous medium replenishment from an external source. Experimentally, this condition is realized in bioreactors such as those used to culture cells at high density for biotechnology purposes (e.g. antibody production, Mercille et al. 2000). Under these conditions, viable cells are expected to reach a steady state given by the dynamic equilibrium between proliferation and death.

Figure 2 shows the result of a simulation campaign carried out with our numerical model. Parameter values were left unchanged with respect to previous simulations, and the only difference is that in the present simulations we consider a continuous flux of fresh medium that replaces culture medium with rates comparable with those used in the experiments with bioreactors. As we see in figure 2, there is a good correspondence between simulations and actual data for different flow rates.

**Table 1** - Estimated morphologic, kinetic and metabolic parameters for a population of dispersed tumor cells and comparison with actual experimental data.

| Parameter | Simulated | | | Experimental | Reference |
|---|---|---|---|---|---|
| *Morphologic* | Average | Min | Max | | |
| Radius (µm) | 5.0 | 4.8 | 5.3 | 5.5 - 7.1 | Freyer and Sutherland 1980 |
| Volume (µm$^3$) | 530 | 471 | 623 | 700 - 1500 | Freyer and Sutherland 1980; Kunz-Schugart et al. 1996 |
| Mitochondria/cell | 220.4 | 190.6 | 266.9 | 83 - 677[a] | Robin and Wong 1988 |
| *Kinetic* | | | | | |
| Growth rate[b] (h$^{-1}$) | 0.035 | | | 0.03 - 0.035[c] | Chignola et al. 2007 |
| Doubling time[b] (h) | 19.8 | | | 19.7 - 22.8[c] | Chignola et al. 2007 |
| G1 (%) | 52.5 | 48.4 | 59.3 | 54.4 ± 2.2[c] | Chignola and Milotti 2005 |
| S (%) | 34.5 | 30.5 | 40.5 | 27.5 ± 5.8[c] | Chignola and Milotti 2005 |
| G2/M (%) | 12.9 | 7.3 | 17.7 | 16.4 ± 1.7[c] | Chignola and Milotti 2005 |
| *Metabolic* | | | | | |
| ATP/Cell[d] | 5.5 | 5.4 | 5.6 | 4.3 - 5.8 | Chignola et al. 2007 |
| Glucose uptake[e] | 1.9 ± 0.3 | | | 2.5 ± 0.2 | Rodriguez-Enriquez et al. 2000 |
| Lactate production[e] | 3.8 ± 0.3 | | | 3.9 ± 0.8 | Rodriguez-Enriquez et al. 2000 |
| ATP production[e,f] | 19.8 ± 8.3 | | | 37.8 | Rodriguez-Enriquez et al. 2000 |
| ATP production[e,g] | 10.6 ± 1.3 | | | 11.4 ± 2.3 | Rodriguez-Enriquez et al. 2000 |
| Oxygen consumption[e] | 0.25 ± 0.1 | | | 0.48 ± 0.1 | Rodriguez-Enriquez et al. 2000 |

[a]Range of the number of mitochondria observed in different cell types.
[b]The growth rate for both simulated and experimental cell populations was calculated by exponential fitting of growth curves. The doubling time was then calculated as log2/(growth rate).
[c]Data measured for MOLT3 (human T lymphoblastoid cell line) and Raji (human B lymphoblastoid cell line) cells in our own experiments.
[d]Values are expressed as 10$^{-18}$ kg.
[e]Values are expressed as 10$^{-19}$ kg s$^{-1}$.
[f]ATP production through oxidative phosphorylation
[g]ATP production through glycolysis

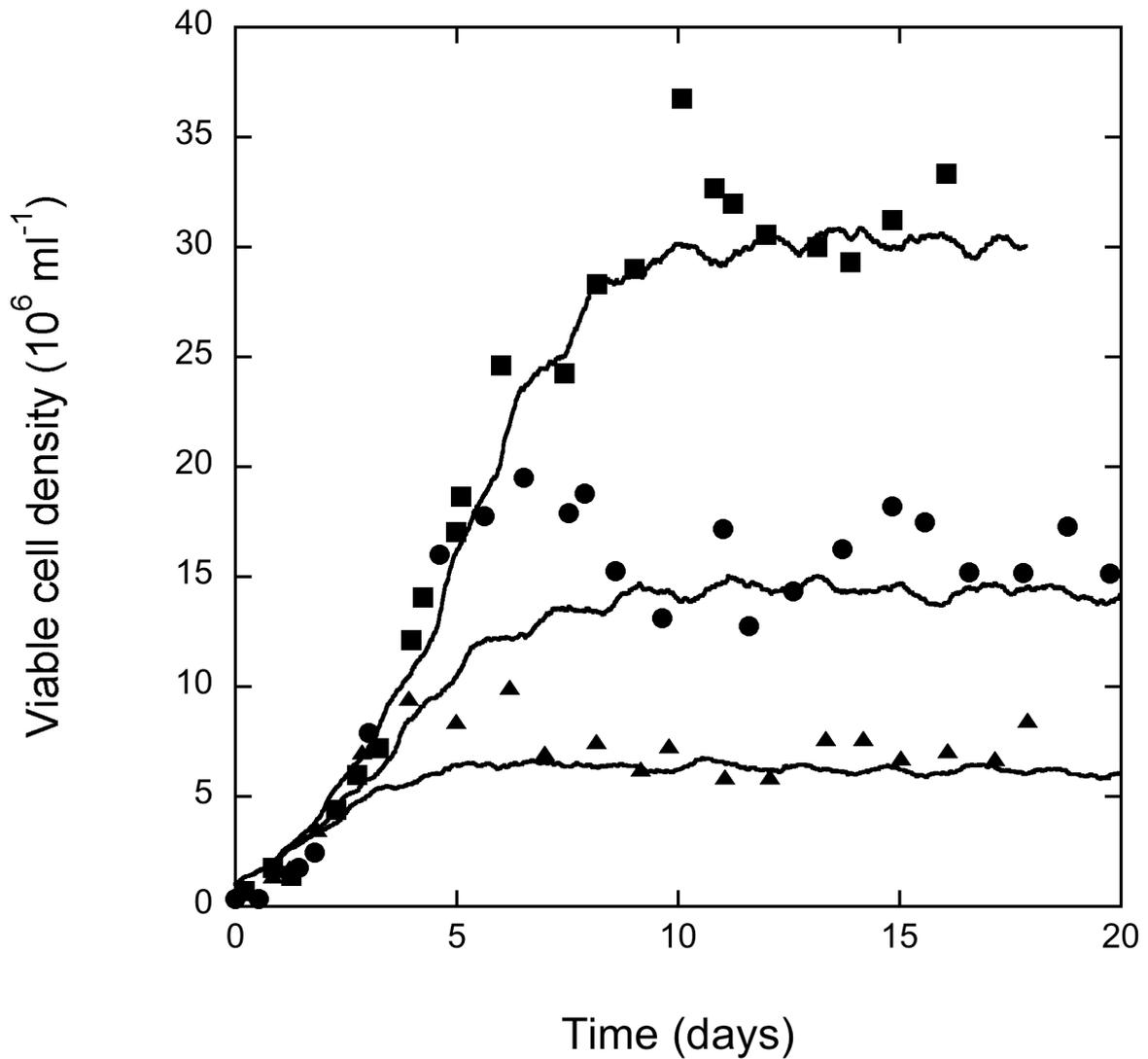

Fig.2 - Growth kinetics of dispersed cells in a bioreactor. Points represent experimental data taken at different flux rates of fresh medium and are: squares, 2.33 volumes of fresh medium pefused / effective suspension volume / day (vvd); circles, 1 vvd; triangles, 0.48 vvd. Lines represent simulation outputs obtained with our program for the same fluxes. See also the text for further details. Experimental data have been drawn from (Mercille at al. 2000).

This kind of virtual experiment was not planned during model development and model parameters were not tuned to take into account the growth of simulated cells in an open environment. This further demonstrates the predictive power of the model and its potential use as an *in silico* cell biology laboratory.

*3.3 Climbing the third dimension: a simulation test run of the growth of tumor spheroids*

We have shown above that our model of the cell can be utilized to simulate the behavior of large tumor cell populations growing both in closed or in open environments. The very same model can also be used to simulate the growth of avascularized tumor spheroids, our more ambitious goal. Obviously, this means that we have to take into consideration new biological, chemical, physical and mathematical aspects.

Cells in a spheroid grow attached to each other. Thus, one must include and model biomechanical forces that act upon cells. Cells have been modeled as soft spheres interacting through visco-elastic forces (Chignola and Milotti 2004). These forces allow the whole cell aggregate to preserve a three-dimensional structure in spite of the repulsion that cells exert at mitosis on neighboring cells to compete for space. This result in a dynamic balance between repulsive and adhesion forces that shape the spheroid structure during its life.

Each cell is surrounded by a small free volume that models the extracellular space. This is fundamental to allow nutrients and waste molecules to diffuse freely through the cell cluster. The inclusion of diffusion, however, introduces processes that occur with very short characteristic times, in the order of a few μs, and this increases the stiffness of the underlying system of differential equations.

Figure 3 shows a result obtained in a preliminary test run with a small spheroid of 1000 cells (the cluster in figure 3 has a diameter of approximately 100 μm). In this case cells are represented by semitrasparent balls so that the figure is actually a 2D projection of the cell cluster. The colors represents the oxygen concentration within the cluster, that is computed along with all the other cellular variables, during the time evolution of the cluster itself. The oxygen concentration decreases towards the center of the spheroid and one can easily detect a spatial gradient that is qualitatively similar to that observed in real experiments. Eventually, as the spheroid grows the internal environment will become hypoxic thus leading to cell death and to the formation of a necrotic core (in this simulation internal cells are still alive). It is very important to remark that at this spatial resolution one can observe anisotropies of the concentration profiles. Since concentration profiles are computed during the time evolution of the whole cluster, the dynamic variation of the distributions of molecular species can also be appreciated and studied for the first time. This computational approach could eventually help to improve our knowledge on the initial growth phase of avascular solid tumors, as it discloses details that are otherwise very difficult to detect and measure.

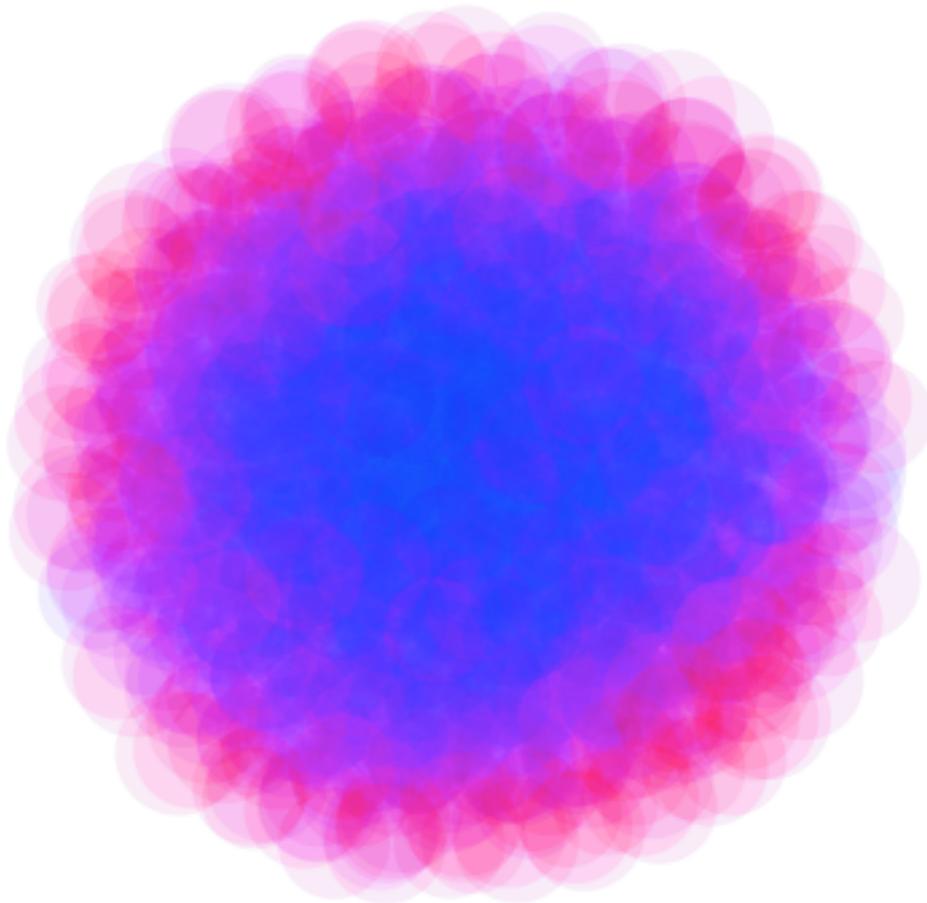

Fig.3 - Oxygen concentration profiles in a small simulated spheroid (approximately 100 μm diameter). Here cells are represented as semitrasparent balls, and oxygen concentration profiles in the inner layers of the spheroid become visible. Oxygen concentration is given in colors: red, higher concentration; blue, lower concentration.

# 4. Discussion and Conclusions

*4.1 Too many equations and parameters?*

We are modeling avascularized tumor spheroids starting from individual molecular reactions in single cells and climbing the ladder of complexity up to the behavior of whole multicellular clusters. As we have shown above, this approach implies that one must deal with very different space and time scales, with the latter spanning 12 orders of magnitude. This effort unavoidably translates into a complex mathematical desciption that involves a vast number of equations, logical rules and parameters, and with many free parameters one can in principle simulate any number of different patterns. Thus, a central question is whether our model could indeed be considered a reliable description of metabolism and mechanical evolution of cell clusters.

We have followed a conservative approach: on one hand, we extensively searched the scientific literature to find estimates for as many parameters as possible. This is the case of diffusion constants of molecular species and for 22 out of 53 parameters that are included into the core model of cell metabolism and proliferation. On the other hand, many parameters assume values that have been estimated through independent biophysical modeling of experimental data, and among them are all the biomechanical parameters and 10 metabolic parameters. Of the remaining 21 metabolic parameters, 19 were estimated by fitting the whole model to experimental data and 2 assume values whose range was estimated on the basis of biophysical considerations, independent modeling and data fitting. Most importantly, once fixed the parameter values are not changed any further and simulation outputs are compared with new sets of experimental data to test the predictive power of the model itself (as we have shown here for cells growing in a bioreactor). All comparisons between simulation outputs and experimental data are finally carried out on a strict quantitative basis.

Thus we feel confident that our numerical simulator is indeed a reliable model of the growth of tumor cells and tumor cell clusters.

Interestingly, it was found that most parameters are correlated and can by no means assume arbitrary values. This is mainly due to the strong feedback between cells and the environment where simulated cells grow and that we have modeled. For example, one might tune metabolic parameters to allow cells to utilize nutrients more efficiently, produce energy under the form of ATP and grow faster. But this consequently results in a higher production and secretion of waste molecules that increase rapidly the acidity of the medium thus leading to cell death. Many parameters of the metabolic network of the cell are also strongly correlated, and this is a typical consequence of the interconnections between reactions that utilize different substrates for the same purpose, such as in the case of ATP and glutamine for both protein and DNA synthesis. As a consequence, the actual dimension of the parameter space is much smaller than the total number of parameters, and parameter tuning is far less complex than it appears to be at first sight.

*4.2 A modeling exercise?*

The final goal of our effort is the development, step by step, of a numerical tool to simulate realistically the growth of avascular tumors and thus to explore the initial growth phase of solid tumors. This kind of numerical simulation has several important implications which are listed below:

1. it is possible to perform virtual experiments *in silico* that complement *in vitro* measurements, where many parameters are not directly accessible, and also *in vivo* observations, where accessibility problems are even greater also because of ethical issues. Our simulation program is indeed a virtual laboratory where one can make experiments at will in due time, and we hope that in the near future it will drive experimenters towards the search of yet unexplored biological properties of tumors;

2. the simulation focuses the modeling effort on the important details of cellular biophysics and spawns new ideas, both theoretical and experimental. For example, one important aspect that we have investigated to test the validity of our simulation program is whether, and eventually under which conditions, it could simulate a cell population with desynchronized cell cycles as it is observed for real cells (Chignola and Milotti 2005; Chignola et al. 2007). This prompted us to think at the sources of internal randomness in cells, and in the attempt to investigate the biological causes of cell cycle desynchronization we then developed both theoretical tools and carried out new experimental observations (Milotti et al. 2008b; Tomelleri et al. 2008);

3. the numerical model includes many complex non-linear interactions between different parts of the cell, and thus it has interesting predictive properties as unexpected biological behaviors can emerge;

4. the model integrates several parts of our knowledge of cell biology, a knowledge that is fragmented in a huge number of small pieces throughout the scientific literature. An important aspect of our effort is that we are trying to connect together at least part of these pieces, and see whether they can provide a coherent picture. The model produces results that compare favourably with experimental data, and this indicates that it is possible to understand the cells' functions at the systemic level in quantitative terms;

5. because of its incremental structure, our simulation program may serve as a platform to test the validity of other models of specific biochemical circuits.

For all the above reasons, we believe that our effort is by no means just a modeling exercise, but a serious and novel attempt to model cell biology.

**Acknowledgements**

This work is supported by grants from the Istituto Nazionale di Fisica Nucleare, Group V, experiment VBL-Rad.

**References**

Agmon, N. 1995. The Grotthuss mechanism. *Chem Phys. Lett.* 244: 456-62.

Barabasi, A. L., Oltvai, Z. N. 2004. Network biology: understanding the cell's functional organization. *Nat. Rev. Genet.* 5: 101-13.

Binney, J. J., Dowrick, N. J., Fisher, A. J., Newman, M. E. J. 1998. *The theory of critical phenomena*. Oxford: Oxford University Press.

Böckmann, B., Grill, H. J., Giesing, M. 2001. Molecular characterization of minimal residual cancer cells in patients with solid tumors. *Biomol. Eng.* 17: 95-111.

Chignola, R., Milotti, E. 2004. Numerical simulation of tumor spheroid dynamics. *Physica A* 338: 261-6.

Chignola, R., Milotti, E. 2005. A phenomenological approach to the simulation of metabolism and proliferation dynamics of large tumour cell populations. *Phys. Biol.* 2: 8-22.

Chignola, R., Dalla Pellegrina, C., Del Fabbro, A., Milotti, E. 2006. Thresholds, long delays and stability from generalized allosteric effect in protein networks. *Physica A* 371: 463-72.

Chignola, R., Del fabbro, A., Dalla Pellegrina, C., Milotti, E. 2007. Ab initio phenomenological simulation of the growth of large tumor cell populations. *Phys. Biol.* 4: 114-33.

De Berg, M., Van Kreveld, M., Overmars, M., Schwarzkopf, O. 2000. *Computational geometry: algorithms and applications*. New York: Springer-Verlag.

Freyer, J. P., Sutherland, R. M. 1980. Selective dissociation and characterization of cells from different regions of multicell tumor spheroids. *Cancer Res.* 40: 3956-65.

Hanahan, D., Weinberg, R. A. 2000. The hallmark of cancer. *Cell* 100: 57-70.

Köstler, W. J., Brodowicz, T., Hejna, M., Wiltschke, C., Zielinski, C. C. 2000. Detection of minimal residual disease in patients with cancer: a review of techniques, clinical implications, and emerging therapeutic consequences. *Cancer Detect. Prev.* 24: 376-403.

Kunz-Schugart, L. A., Groebe, K., Mueller-Klieser, W. 1996. Three-dimensional cell culture induces novel proliferative and metabolic alterations associated with oncogenic transformation. *Int. J. Cancer* 66: 578-86.

Levintow, L., Eagle, H., Piez, K. A. 1955. The role of glutamine in protein biosynthesis in tissue culture. *J. Biol. Chem.* 215: 441-60.

Mercille, S., Johnson, M., Lanthier, S., Kamen, A. A., Massie, B. 2000. Understanding factors that limit the productivity of suspension-based perfusion cultures operated at high medium renewal rates. *Biotechnol. Bioeng.* 67: 435-50.

Mueller-Klieser, W. 1997, Three-dimensional cell cultures: from molecular mechanisms to clinical applications. *Am. J. Physiol.* 273 (*Cell Physiol.* 42): C1109-C1123.

Milotti, E., Del Fabbro, A., Dalla Pellegrina, C., Chignola, R. 2007. Dynamics of allosteric action in multisite protein modification. *Physica A* 379: 133-50.

Milotti, E., Chignola, R., Dalla Pellegrina, C., Del Fabbro, A., Farina, M., Liberati, D. 2008a. VBL: Virtual biophysics Lab. *Nuovo Cimento* 31C: 109-18.

Milotti, E., Del Fabbro, A., Dalla Pellegrina, C., Chignola, R. 2008b. Statistical approach to the analysis of cell desynchronization data. *Physica A* 387: 4204-14.

Milotti, E., Del Fabbro, A., Chignola, R. 2009. Numerical integration methods for large-scale biophysical simulations. *Comput. Phys. Commun.* in press (doi: 10.1016/j.cpc.2009.06.021).

Nelson, D. L., Cox, M. M. 2005. *Lehninger principles of biochemistry*. New York: Freeman.

Phillips, J. C., Zheng, G., Kurmar, S., Kal, L. V. 2002. NAMD: biomolecular simulation on thousands of processor. Proceedings of the IEE/ACM SC2002 Conference, Technical Paper 277, IEEE Press, New York.

Robin, E. D., Wong, R. 1988. Mitochondria DNA molecules and virtual number of mitochondria per cell in mammalian cells. *J. Cell Physiol.* 136: 507-13.

Rodriguez-Enriquez, S., Torres-Marquez, M. E., Moreno-Sanchez, R. 2000. Substrate oxidation and ATP supply in AS-300 hepatoma cells. *Arch. Biochem. Biophys.* 375: 21-30.

Salzman, N. P., Eagle, H., Sebring, E. D. 1958. The utilization of glutamine, glutamic acid, and ammonia for the biosynthesis of nucleic acid bases in mammalian cell cultures. *J. Biol. Chem.* 230: 1001-12.

Sutherland, R. 1988. Cell and environment interactions in tumor microregions: the multicell spheroid model. *Science* 240: 177-84.

Tomelleri, C., Milotti, E., Dalla Pellegrina, C., et al. 2008. A quantitative study of the growth variability of tumour cell clones in vitro. *Cell Prolif.* 41: 177-91.

Vajkoczy, P., Farhadi, M., Gaumann, A., et al. 2002. Microtumor growth initiates angiogenic sprouting with simultaneous expression of VEGF, VEGF receptor-2, and angiopoietin-2. *J. Clin. Invest.* 109: 777-85.

Weinberg, R. A. 1995. The retinoblastoma protein and cell cycle control. *Cell* 81: 323-30.